\documentclass[11pt]{article}
\usepackage{epsfig}
\usepackage{latexsym}
\usepackage{amsmath}

\newtheorem{theorem}{Theorem}{\bf}{\it}
\newtheorem{definition}[theorem]{Definition}{\bf}{\rm}
{\bf}{\it}
{\bf}{\it}
\newtheorem{lemma}[theorem]{Lemma}{\bf}{\it}
          \def\dt{\cal}

          \def\dM{{\dt M}}
          \def\dN{{\dt N}}

          \def\E{{\cal E}}

          \def\H{{\cal H}}

          \def\U{{\cal U}}

          \def\gO{\Omega}
          
          \def\go{\omega}

      \def\h{{\bf h}}
          \def\Halmos{\quad\hfill$\Box$}

          \def\naturals{{\bf N}}

          \def\reals{{\bf R}}

\title{$\beta$-Boundedness, Semipassivity, and the KMS-Condition}
\author{Bernd Kuckert\\University of Amsterdam,
Korteweg-de Vries Institute for Mathematics\thanks{On leave for:
II. Institut f\"ur Theoretische Physik, Luruper Chaussee 149, 22761 Hamburg,
Germany, e-mail: bernd.kuckert@desy.de}\\Plantage Muidergracht 24,
1018 TV Amsterdam, The Netherlands\\e-mail: kuckert@science.uva.nl}
\date{November 2001}

\begin{document}
\maketitle

\begin{abstract}
The proof of
a recent result by Guido and Longo establishing the equivalence
of the KMS-condition with complete $\beta$-boundedness \cite{GL01}
is shortcut and generalized in such a way that a
covariant version of the theorem is obtained.
\end{abstract}

Recently it was proved by Guido and Longo in \cite{GL01} that the
KMS-condition at a finite nonnegative temperature is equivalent
to a condition called {\it complete $\beta$-boundedness}, which
imposes a bound on the number of degrees of freedom in certain
phase space regions and is a weak form of the Buchholz-Wichmann
nuclearity condition \cite{BuWi}, very similar to the (weaker)
Haag-Swieca compactness criterion \cite{HaSw}. On the other hand,
it was shown in \cite{Kuc01} that the KMS-condition at nonnegative
temperature in {\it some} (a priori unknown) inertial frame is
equivalent to a condition called {\it complete semipassivity}.

Both proofs are variations
of the classical result by Pusz and Woronowicz \cite{PW78}, who proved that
the KMS-condition at nonnegative temperature
is equivalent to complete passivity.

It is of interest to investigate bounds on
the efficiency of thermodynamic cycles in less generic settings
than that of a stationary and homogeneous state, as the extent
to which the passivity condition is violated can be considered as
a kind of a nonequilibrium state's
distance from thermodynamic equilibrium.

As a first step on this path, the result
found by Guido and Longo is generalized below in such a way that it
characterizes semipassive states as well. As a spinoff of this
result, a shortcut of the Guido-Longo argument is given first.

In what follows, the algebra of observables of the system under
consideration is a von Neumann algebra $\dM$ on a Hilbert space
$\H$, and the state $\go$ of $\dM$ under consideration is induced by
a cyclic unit vector $\gO$. The time evolution is generated by a
selfadjoint operator $H$ with the property that $e^{itH}\dM e^{-itH}=\dM$
and $H\gO=0$. We fix a parameter $\beta\geq0$ that will, eventually,
estimate the inverse temperature of $\go$.

\begin{definition}
The state $\go$ is called {\bf $\beta$-bounded with bound 1}
if the linear space
$\dM\gO$ is a subspace of the domain of $e^{-\beta H}$ and
if the set $e^{-\beta H}\dM_1\gO$ consists of vectors with lengths $\leq1$
where $\dM_1:=\{A\in\dM:\,\|A\|\leq1\}$.

$\go$ is called {\bf completely $\beta$-bounded} if
for each $n\in\naturals$,
the state $\go^{\otimes n}$ on the algebra $\dM\otimes\dots\otimes\dM$
is $\beta$-bounded with bound 1.
\end{definition}

\begin{theorem}[Guido, Longo]\label{X}
$\go$ is completely $\beta$-bounded if and only if it is a ground state
or a KMS-state at an inverse temperature $\geq2\beta$.
\end{theorem}
{\it Proof.} The condition is sufficient by Lemma 1.2 in \cite{GL01}.
It remains to prove that it is necessary.

If $E$ denotes the orthogonal projection onto the closure of the
subspace $\h:=\overline{\dM'\gO}$ and if $\Delta$ denotes the modular
operator of $\gO$ with respect to
the von Neumann algebra $\dN$ obtained by
restricting the elements of $\dM$ to $\h$ (note that
$\gO$ is not only cyclic, but also separating with respect to this
von Neumann algebra on $\h$), then $\beta$-boundedness of $\go$ implies,
by Cor. 1.8 in \cite{GL01}, that
$$e^{-2\beta H}\leq 1+\Delta E.$$

If $\go$ is faithful, then this implies that each point $(\eta,\kappa)$
in the joint spectrum $\sigma_{H,K}$
of the operators $H$ and $K:=\log(\Delta)$ satisfies
\begin{equation}\label{constraint}
e^{-2\beta\eta}\leq 1+e^{\kappa},
\end{equation}
and as $\go$ is completely $\beta$-bounded, it follows
by the arguments used in \cite{PW78} that this
inequality must hold for the elements of the additive
group generated by the elements of $\sigma_{H,K}$ as
well.

Ineq. (\ref{constraint}) can be rewritten as
%\begin{equation}\label{constraintb}
$$\eta\geq-\frac{1}{2\beta}\log(1+e^{\kappa})=:f(\kappa),$$
%\end{equation}
and this estimate separates $\sigma_{H,K}$
from a region containing an open cone. Each subgroup in the admitted
region must, therefore, be a subgroup of a one-dimensional
subspace, and in particular, $\sigma_{H,K}$ and the additive subgroup
it generates must be subsets of such a one-dimensional space $X$.
As $f(\kappa)$ is defined everywhere, this subspace
cannot be the $\eta$-axis, so $H=-\alpha K$ for some $\alpha\in\reals$.

Since $f(\kappa)$ tends to zero as $\kappa\to-\infty$, $\alpha$
must be nonnegative, and since $f(\kappa)$ tends to
$-\frac{\kappa}{2\beta}$ as $\kappa\to+\infty$, it follows that
$\alpha\leq1/(2\beta)$, so $\go$ is a KMS-state at an inverse
temperature $\geq 2\beta$ if $\alpha\neq0$, and if $\alpha=0$,
then $\go$ trivially is a ground state of $H=0$.

It remains to consider the case that $\go$ is not faithful. Then the elements
of the space $\h^\perp=(1-E)\H$ are eigenvectors of $\Delta E$ with the
eigenvalue zero. As $\dM'$ is invariant under the adjoint action of
$e^{itH}$, the space $\h$ is invariant under $e^{itH}$, so
$H|_{\h^\perp}$ is a self-adjoint
operator in $\h^\perp$, whose spectral projections are
restrictions of the corresponding spectral projections of
$H$, respectively.
But this implies that $\sigma_{H,\Delta E}$
contains some point of the form $(\eta,0)$.

Now let $(\eta',\delta')$ be an arbitrary point in
$\sigma_{H,\Delta E}$. As $\go$ satisfies complete
$\beta$-boundedness, each point of the form
$(n\eta'+\eta,n\delta'\cdot0)$, $n\in\naturals$, must satisfy
Eq. (\ref{constraint}) as well (cf. \cite{PW78,Kuc01}), so
$2\beta(n\eta'+\eta)\geq0$ for all $n\in\naturals$,
which can be only if $\eta'\geq0$.

So we have proved that $\eta'\geq0$ for all
$(\eta',\delta')\in\sigma_{H,\Delta E}$, which implies $H\geq0$
(cf. Lemma B.2 in \cite{Kuc01}).

\Halmos

The above proof does not only provide a more direct argument than the
original proof, it also admits a generalization of the theorem
in the spirit of Thm. 3.3 in \cite{Kuc01}.

To this end, assume that $H$ generates, together with $s$
self-adjoint operators $P_1,\dots,P_s$ collected in a vector operator
${\bf P}$, a strongly continuous representation of the $1+s$-dimensional
spacetime translation group that leaves the vector $\gO$ invariant.
$\go$ is stationary in all inertial frames, whereas in the presence
of matter, it can be a thermodynamic equilibrium state in that
matter's rest frame only. The reason is that the condition of
passivity, which is a consequence of the Second Law, is violated
in the other frames.

The appropriate weakening of the passivity condition is
semipassivity. In the following definition, $\U_1(\dM)$
denotes the group of all unitary operators in $\dM$ that
can be connected to the unit operator by a norm-continuous
path of unitary operators in $\dM$.

\begin{definition}
$\go$ is called {\bf semipassive} if there exists a constant $\E\geq0$
such that
$$-\langle W\gO,HW\gO\rangle\leq\E\langle W\gO,|{\bf P}| W\gO\rangle$$
for each $W\in\U_1(\dM)$ with $[H,W]\in\dM$ and
$[P_1,W],\dots,[P_s,W]\in\dM$.\footnote{
As a quadratic form, the commutator of $H$ and
$W$ is defined on the domain of $H$. The condition $[H,W]\in\dM$
means that this quadratic form is bounded and that its associated
bounded operator is an element of $\dM$. $[P_1,W],\dots,
[P_s,W]\in\dM$ is to be read accordingly.}
Each constant $\E$ satisfying this condition
is called an {\bf efficiency bound} of $\go$.
$\go$ is called {\bf completely semipassive} if all its tensorial
powers are semipassive with respect to the same efficiency bound.
\end{definition}

We recall Thm. 3.3 from \cite{Kuc01}:

\begin{theorem}\label{Hauptsatz}
$\go$ is completely semipassive with efficiency bound $\E\geq0$
if and only if there exists a ${\bf u}\in\reals^s$ with
$|{\bf u}|\leq\E$ such that $\go$ is a ground state
or a KMS-state (at finite
$\beta\geq0$) with respect to $H+{\bf uP}$.
\end{theorem}

This result describes a most generic example of a nonequilibrium
state, and the question is whether bounds on the power of a
cyclic process could be of interest in less generic situations.
As far as such investigations are concerned, it is an obstacle of
the above definition of semipassivity that the invariance of
$\go$ is part of the definition and its motivation. While the
problem addressed in Thm. \ref{Hauptsatz} is nontrivial only if
$\go$ is invariant under all spacetime translations, it would be
of interest whether the semipassivity condition can be subdivided
into this invariance property plus some additional condition that
may be meaningful in other situations as well. Such a condition
is semi-$\beta$-boundedness.

\begin{definition}
The state $\go$ is called {\bf semi-$\beta$-bounded} if there
exists a damping factor $\E\geq0$ such that the linear space
$\dM\gO$ is a subspace of the domain of $e^{-\beta(H+\E|{\bf
P}|)}$ and the set $e^{-\beta(H+\E|{\bf P}|)}\dM_1\gO$ consists
of vectors with length $\leq1$.

It is called {\bf completely semi-$\beta$-bounded} if for each
$n\in\naturals$, the state $\go^{\otimes n}$ on the algebra
$\dM\otimes\dots\otimes\dM$ is semi-$\beta$-bounded with respect
to one fixed damping factor $\E\geq0$.
\end{definition}
As $|{\bf P}|$ is a positive operator, the operator
$e^{-\beta\E|{\bf P}|}$ is bounded and provides an additional
damping term, so $\beta$-boundedness implies
semi-$\beta$-boundedness for all $\E\geq0$, i.e.,
semi-$\beta$-boundedness is the weaker assumption.

\begin{theorem}\label{main}
A stationary and homogeneous state $\go$ is completely
semi-$\beta$-bounded
if and only if there exists a ${\bf u}\in\reals^s$ with $|{\bf
u}|\leq\E$ such that $\go$ is a ground state or a KMS-state at an
inverse temperature $\geq 2\beta$ with respect to $H+{\bf uP}$.
\end{theorem}
{\it Proof.} As in the proof of Thm. \ref{X}, the only nontrivial part
is the proof that the condition is necessary.

One checks that the proof of Cor. 1.8 in
\cite{GL01} still works if one replaces $H$ by $H+\E|{\bf P}|$,
and one obtains
\begin{equation}\label{Ungleichung}
e^{-2\beta(H+\E|{\bf P}|)}\leq 1+\Delta E,
\end{equation}
where $\Delta$ and $E$ are defined as in the proof of
Thm. \ref{X}. Again, we distinguish the cases that
$\go$ is faithful and not faithful:

\begin{lemma}\label{lemma1}
If $\go$ is faithful, then there exists a ${\bf u}\in\reals^s$
such that either

(i) $H+{\bf uP}=0$, or

(ii) $\go$ is a KMS-state at an inverse temperature
$\geq2\beta$ with respect to $H+{\bf uP}$.
\end{lemma}
{\it Proof.}
Ineq. (\ref{Ungleichung}) implies that
for each $(\eta,{\bf k},\kappa)\in\sigma_{H,{\bf P},K}$, one has
\begin{equation}\label{Ungleichung2}
\eta+\E|{\bf k}|\geq-\frac{1}{2\beta}\ln(1+e^{\kappa}),
\end{equation}
which expells the joint spectrum of $H$, ${\bf P}$, and $K$
from a region containing an open cone. As in \cite{PW78}
it follows that
$\sigma_{H,{\bf P},K}$ is a subset of a subspace $X$ of
$\reals^{s+2}$ with codimension $\geq1$ whose elements
satisfy Ineq. (\ref{Ungleichung2}).

If $H$ is a linear function of ${\bf P}$, then there exists
a ${\bf u}\in\reals^s$ such that $H+{\bf uP}=0$. Inserting
this into Ineq. \ref{Ungleichung2}, one finds that
$$-{\bf uk}+\E|{\bf k}|\geq\frac{1}{2\beta}\ln(1+e^{\kappa})$$
whenever the pair $({\bf k},\kappa)$ is in the joint
spectrum of ${\bf P}$ and $K$, and by complete semiboundedness,
one also obtains
\begin{equation}\label{Ungleichung3}
-n{\bf uk}+n\E|{\bf k}|\geq\frac{1}{2\beta}\ln(1+e^{n\kappa})
\end{equation}
for all $n\in\naturals$. If $J$ denotes the modular conjugation
associated with $\dM$ and $\gO$, then $JHJ=-H$, $J{\bf P}J=-{\bf P}$,
and $JKJ=-K$ by elementary Tomita-Takesaki theory,
so Ineq. (\ref{Ungleichung3}) holds for all
$n\in{\bf Z}$. But this immediately entails $|{\bf u}|\leq\E$,
proving Alternative (i) of the statement.

There remains the case that $H$ is not a linear function of ${\bf P}$.

As $X$ cannot contain the $\kappa$-axis by Ineq. (\ref{Ungleichung2}),
$K$ is a linear function of $H$ and ${\bf P}$, so there exists an
$\alpha\in\reals$ and a ${\bf v}\in\reals^s$ such that
\begin{equation}\label{KundHundP}
K=-\alpha H+{\bf vP}.
\end{equation}
The vector ${\bf v}$ is unique up to a component that is perpendicular
to the smallest linear subspace $Y$ of $\reals^s$ containing the joint
spectrum of the components of ${\bf P}$, so ${\bf v}$ can be chosen in
$Y$.

If ${\bf vP}=0$, then $K=-\alpha H$, and Ineq. (\ref{Ungleichung2})
implies that $\alpha\leq2\beta$, so $H$ generates a KMS-dynamics
at an inverse temperature $\geq2\beta$.

In the remaining case that ${\bf vP}\not=0$, the unit vector ${\bf
e_v}:=|{\bf v}|^{-1}{\bf v}$ is in $Y$.

If $\alpha\not=0$, then Eq. (\ref{KundHundP}) and the assumption that
$H$ is not a function of ${\bf P}$ entail $K\not=0$, so for each
$\kappa>0$ and each $\lambda>0$, one has $(\eta(\lambda,\kappa),
\lambda{\bf e_v},\kappa)\in X$, where $$\eta(\lambda,\kappa)
:=-\frac{1}{\alpha}(\kappa -\lambda{\bf e_v v})
=-\frac{1}{\alpha}(\kappa-\lambda|{\bf v}|).$$
Ineq. (\ref{Ungleichung2}) yields
$$-\frac{\kappa}{\alpha}-
\lambda\left(\left|\frac{{\bf v}}{\alpha}\right|-\E\right)
\geq-\frac{1}{2\beta}\ln(1+e^{\kappa})$$
for all $\kappa,\lambda>0$, so $\alpha\leq2\beta$ and
$\left|\frac{{\bf v}}{\alpha}\right|\leq\E$,
and defining ${\bf u}:=\frac{{\bf v}}{\alpha}$, one finds that $H+{\bf uP}$
generates a KMS-dynamics at an inverse temperature $\geq2\beta$, and one
arrives at Alternative (ii) of the statement.

It remains to consider the case that $\alpha=0$, i.e., that $K={\bf v
P}$.  As $H$ is not a linear function of ${\bf P}$, while $K={\bf
vP}$, $H$ cannot be a linear function of $K$ and ${\bf P}$, so $X$
must contain the $\eta$-axis. But if $(\eta,{\bf k},\kappa)\in
\sigma_{H,{\bf P},K}$,
then so is $(\eta',{\bf k},\kappa)$ for all
$\eta'\in\reals$, and Ineq. (\ref{Ungleichung2}) implies
$$\eta'+\E|{\bf k}|\geq-\frac{1}{2\beta}\ln(1+e^{\bf vk})$$ for all
$\eta'\in\reals$, which cannot be.

\Halmos

\begin{lemma}\label{lemma2}
If $\go$ is not faithful, then there exists a ${\bf u}\in\reals^s$
with $|{\bf u}|\leq\E$
such that $\go$ is a ground state with respect to $H+{\bf uP}$.
\end{lemma}
{\it Proof.} The elements
of the space $\h^\perp=(1-E)\H$ are eigenvectors of $\Delta E$ with the
eigenvalue zero. As $\dM'$ is invariant under the adjoint action of
$e^{itH}$, the space $\h$ is invariant under $e^{itH}$ and
$e^{itP_1},\dots,e^{itP_s}$, so
$H|_{\h^\perp}$ and $P_1|_{\h^\perp},\dots,P_s|_{\h^\perp}$
are self-adjoint
operators in $\h^\perp$, whose spectral projections are
restrictions of the corresponding spectral projections of
$H$ and ${\bf P}$, respectively.
But this implies that $\sigma_{H,{\bf P},\Delta E}$
contains some point of the form $(\eta,{\bf k},0)$.

Now let $(\eta',{\bf k}',\delta')$ be an arbitrary point in
$\sigma_{H,{\bf P},\Delta E}$. As $\go$ satisfies complete
semi-$\beta$-boundedness, each point of the form
$(\eta+n\eta',{\bf k}+n{\bf k}', 0\cdot n\delta')$,
$n\in\naturals$, must satisfy Ineq. (\ref{Ungleichung2}) as well,
so $\eta+n\eta'+\E|{\bf k}+n{\bf k}'|\geq0$ for all
$n\in\naturals$, which can be only if $\eta'\geq\E|{\bf k'}|$.

So we have proved that
\begin{equation}\label{Ungleichung4}
\eta'\geq\E|{\bf k'}|
\end{equation}
for all $(\eta',{\bf k'},\delta')\in\sigma_{H,{\bf P},\Delta E}$
and, hence, for all $(\eta',{\bf k}')\in\sigma_{H,{\bf P}}$ (use,
e.g., Lemma B.2 in \cite{Kuc01}). By complete
semi-$\beta$-boundedness, the corresponding estimate should hold
for all tensorial powers of $\go$ as well, which implies that the
joint spectrum of $H$ and ${\bf P}$ is a subset of a
sub-semigroup of $\reals^{1+s}$ all of whose elements satisfy
Ineq. (\ref{Ungleichung4}). Such a semigroup must be a subset of
a half space all of whose elements satisfy Ineq.
(\ref{Ungleichung4}). This implies that there exists a ${\bf
u}\in\reals^s$ with $|{\bf u}|\leq\E$ such that $H+{\bf uP}$ is a
positive operator (cf. Lemma B.2 in \cite{Kuc01}).

\Halmos

With these two lemmas, the proof of Thm. \ref{main} is complete
as well.

\Halmos

\section*{Acknowledgements}
The author thanks D. Arlt and D. Guido for critically reading the 
manuscript.

This work has been supported by the Stichting Fundamenteel Onderzoek 
der Materie (FOM).


\begin{thebibliography}{*****}

\bibitem{BuWi} Buchholz, D., Wichmann, E. H.: Causal Independence
and the Energy-Level Density of States in Local Quantum Field Theory,
{\it Commun. Math. Phys.} {\bf 106}, 321-344 (1986)

\bibitem{GL01} Guido, D., Longo, R.: Natural Energy Bounds in Quantum
Thermodynamics, {\it Commun. Math. Phys.} {\bf 218}, 513-536 (2001)

\bibitem{HaSw} Haag, R. Swieca, J. A.: When Does a Quantum Field Theory 
Describe Particles, {\it Commun. Math. Phys.} {\bf 1}, 308-320 (1965)

\bibitem{Kuc01} Kuckert, B.: Covariant Thermodynamics of Quantum Systems,
preprint {\tt hep-th/0107236}, to appear in {\it Ann. Phys. (N. Y.)}

\bibitem{PW78} Pusz, W., Woronowicz, S. L.: Passive States and KMS States for
General Quantum Systems, {\it Commun. Math. Phys.} {\bf 58}, 273-290 (1978)

\end{thebibliography}
\end{document}